\documentclass{ieeeaccess}
\usepackage{cite}
\usepackage{amsmath,amssymb,amsfonts}
\usepackage{algorithmic}
\usepackage{graphicx}
\usepackage{textcomp}
\usepackage{tabularx}
\usepackage{cite}
\usepackage{float}
\usepackage{caption}
\usepackage{subcaption}
\usepackage{amsmath,amssymb,amsfonts}
\usepackage{hhline}
\usepackage{algorithmic}
\usepackage{graphicx}
\usepackage{dirtytalk}
\usepackage{textcomp}
\usepackage{hyperref}
\usepackage[utf8]{inputenc}
\usepackage{xcolor}
\usepackage{colortbl}
\usepackage[normalem]{ulem}
\useunder{\uline}{\ul}{}
\usepackage{booktabs}
\usepackage{multirow}

\usepackage{bm}
\makeatletter
\AtBeginDocument{\DeclareMathVersion{bold}
\SetSymbolFont{operators}{bold}{T1}{times}{b}{n}
\SetSymbolFont{NewLetters}{bold}{T1}{times}{b}{it}
\SetMathAlphabet{\mathrm}{bold}{T1}{times}{b}{n}
\SetMathAlphabet{\mathit}{bold}{T1}{times}{b}{it}
\SetMathAlphabet{\mathbf}{bold}{T1}{times}{b}{n}
\SetMathAlphabet{\mathtt}{bold}{OT1}{pcr}{b}{n}
\SetSymbolFont{symbols}{bold}{OMS}{cmsy}{b}{n}
\renewcommand\boldmath{\@nomath\boldmath\mathversion{bold}}}
\makeatother

\def\BibTeX{{\rm B\kern-.05em{\sc i\kern-.025em b}\kern-.08em
    T\kern-.1667em\lower.7ex\hbox{E}\kern-.125emX}}

\begin{document}
\doi{}

\title{Democratizing AI: A Comparative Study in Deep Learning Efficiency and Future Trends in Computational Processing}
\author{
Lisan Al~Amin\authorrefmark{1}, 
Md Ismail~Hossain\authorrefmark{2}, 
Rupak~Kumar~Das\authorrefmark{3},
~Mahbubul~Islam\authorrefmark{4},
Abdulaziz Tabbakh\authorrefmark{5},
}

\address[1]{ Cognitive Links,
Maryland, USA (e-mail: alamin@cognitivelinks.llc)}
\address[2]{North South University, Bangladesh}
\address[3]{Pennsylvania State University, USA}
\address[4]{United International University, Bangladesh}
\address[5]{King Fahd University of Petroleum and Minerals (KFUPM)}


\corresp{Corresponding author: Abdulaziz Tabbakh (e-mail: atabakh@kfupm.edu.sa).}

\begin{abstract}
The exponential growth in data has intensified the demand for computational power to train large-scale deep learning models. 
However, the rapid growth in model size and 
complexity raises concerns about equal and fair access to computational resources, particularly under increasing energy and infrastructure constraints.
GPUs have emerged as essential for accelerating such workloads. This study benchmarks four deep learning models (Conv6, VGG16, ResNet18, CycleGAN) using TensorFlow and PyTorch on Intel Xeon CPUs and NVIDIA Tesla T4 GPUs. 
Our experiments demonstrate that, on average, GPU training achieves speedups ranging from 11× to 246× depending on model complexity, with lightweight models (Conv6) showing the highest acceleration (246×), mid-sized models (VGG16, ResNet18) achieving 51-116× speedups, and complex generative models (CycleGAN) reaching 11× improvements compared to CPU training.
Additionally, in our PyTorch vs. TensorFlow comparison, we observed that TensorFlow’s kernel‐fusion optimizations reduce inference latency by approximately 15\%.
We also analyze GPU memory usage trends and projecting requirements through 2025 using polynomial regression. 
Our findings highlight that while GPUs are essential for sustaining AI’s growth, democratized and shared access to GPU resources is critical for enabling research innovation across institutions with limited computational budgets.
\end{abstract}


\titlepgskip=-21pt

\maketitle

\section{Introduction}
\label{sec:introduction}
\PARstart{G}{raphics} Processing Units (GPUs) have transitioned from specialized hardware for rendering graphics to a core component of high-performance computing. Their architectural design—featuring massive parallelism, high memory bandwidth, and exceptional power efficiency (measured in FLOPS per watt)—makes them particularly suitable for computationally intensive tasks~\cite{b1}. This shift has made GPUs indispensable in areas such as scientific computing, artificial intelligence, and deep learning.

Unlike Central Processing Units (CPUs), which excel in sequential task execution, GPUs implement a Single Instruction Multiple Threads (SIMT) model that enables the concurrent execution of thousands of lightweight threads~\cite{b3}. These threads operate across hundreds of simple cores in a lockstep fashion, yielding higher throughput with minimal context-switching overhead. While parallelism on general-purpose CPUs often requires complex programming and is bounded by hardware limitations, GPUs provide a more scalable and hardware-optimized alternative for data-parallel operations such as matrix multiplications, vector computations, and large-scale training.

As data continues to grow exponentially, modern deep learning models have scaled in complexity and size. For example, VGGNet consists of 19 layers and over 144 million parameters~\cite{v10}, resulting in significant computational and memory demands. Training such models on large datasets like CIFAR~\cite{b37}, MS COCO~\cite{b54}, and ImageNet~\cite{b55} can take thousands of iterations involving forward and backward propagation, thus requiring substantial time and hardware resources~\cite{b48}. While multi-core CPUs are accessible, they struggle with such tasks due to limited memory bandwidth and lower degrees of parallelism. In contrast, GPUs can accelerate deep learning pipelines and significantly reduce training and inference time~\cite{b4,b47}.

Several studies demonstrate the superiority of GPUs over CPUs in model training and real-time inference tasks. Convolutional Neural Networks (CNNs), for instance, have been shown to train considerably faster on GPUs~\cite{b5}. Similarly, machine learning algorithms implemented with GPU support report notable improvements in runtime efficiency~\cite{b6}. As the demand for faster model development and deployment grows, the need for efficient GPU-based acceleration becomes increasingly evident. This trend is further supported by the increasing availability of cloud-based GPU resources, such as Google Colab, which aim to democratize access despite constraints on free usage.

Recent advances in AI and machine learning have been accompanied by rapid growth in model size, architectural complexity, and training data requirements\cite{llmssurvey}. Contemporary models increasingly demand vast computational power and memory capacity, making their training feasible primarily on specialized hardware such as GPUs~\cite{aimlcompute}. While GPUs demonstrably outperform CPUs in both training and inference, this performance advantage comes at the cost of increased dependence on scarce computational and energy resources. The global GPU shortage, rising energy consumption, and constraints on data-center expansion pose significant barriers to scaling AI research and industrial adaptation. These challenges disproportionately affect researchers at small institutions, startups, and organizations in developing regions. Consequently, democratizing access to GPU resources through shared, efficient, and energy-aware utilization is desirable and necessary to sustain equitable AI advancement.

In this work, we conduct a detailed performance profiling and analysis comparing CPU and GPU capabilities in deep learning frameworks. While previous studies tend to focus on benchmarking a single framework or hardware configuration, our work contrasts CPU vs. GPU performance across both TensorFlow and PyTorch. We evaluate training and inference runtimes, identifying framework-specific optimizations and hardware utilization patterns. Additionally, we perform a polynomial regression-based trend analysis on GPU memory usage over the past decade to project future hardware requirements. This is a critical step in understanding scaling demands and planning future infrastructure for AI workloads. Furthermore, we contextualize GPU acceleration within the broader challenge of computational resource scarcity, arguing for democratized access to shared GPU infrastructures to support sustainable and inclusive AI research.

Overall, this study aims to quantify the practical advantages of GPUs in deep learning tasks, provide insight into framework-dependent behaviors, and assess how GPU memory trends may influence model development. Our results offer actionable recommendations for practitioners, researchers, and hardware developers working at the intersection of AI and high-performance computing.

The paper is structured as follows: Section II introduces GPU architecture; Section III covers background and related work. Section IV explains the experimental methodology. Section V presents performance results for CPU and GPU comparisons. Section VI analyzes GPU memory trends. Section VII offers discussions, and Section VIII concludes the paper.

\begin{table*}[!htbp]
\caption{Overview of Key Research Contributions in GPU Applications across Diverse Computing Domains}
\centering
\begin{tabular}{|l|l|l|l|l|}
\hline
\begin{tabular}[c]{@{}l@{}}\textbf{Sl} \\ \textbf{No}.\end{tabular} &
\textbf{Category} &
\multicolumn{1}{c|}{\textbf{Authors}} &
\multicolumn{1}{c|}{\textbf{Contributions}} &
\multicolumn{1}{c|}{\textbf{Remarks}} \\ \hline
1 &
\begin{tabular}[c]{@{}l@{}}Parallel Computing\\ \& Machine Learning.\end{tabular} &
Matloff et al \cite{b8} &
Perform parallel computer software &
\begin{tabular}[c]{@{}l@{}}In this paper, customized set up and programmable\\ parallel software concept is used\end{tabular} \\ \hline
2 &
\begin{tabular}[c]{@{}l@{}}GPU \& Machine\\ Learning\end{tabular} &
Steinkraus et al \cite{b13} &
Usage of GPU for Machine Learning &
\begin{tabular}[c]{@{}l@{}}The author proposes a generic two-layer neural\\ network GPU implementation\end{tabular} \\ \hline
3 &
\begin{tabular}[c]{@{}l@{}}GPU \& Deep\\ Learning\end{tabular} &
Gawehn et al \cite{b16} &
GPUs for Deep Learning &
\begin{tabular}[c]{@{}l@{}}Advance drug development by GPU-based\\ learning\end{tabular} \\ \hline
4 &
\begin{tabular}[c]{@{}l@{}}GPU \& Cloud \\ Computing\end{tabular} &
Sisyukov et al \cite{b18} &
\begin{tabular}[c]{@{}l@{}}GPU optimized data analysis of\\ cloud infrastructure\end{tabular} &
\begin{tabular}[c]{@{}l@{}}This paper suggests an approach to organize a\\ private cloud environment for real-time\\ data processing\end{tabular} \\ \hline
5 &
GPU \& Statistics &
Eller et al \cite{b19} &
\begin{tabular}[c]{@{}l@{}}GPU backed massive volume of data\\ analysis\end{tabular} &
\begin{tabular}[c]{@{}l@{}}Perform high statistical GPU analyzes to explore \\ fundamental neutrino properties\end{tabular} \\ \hline
6 &
GPU \& Data Analytics &
Nie et al \cite{b21} &
\begin{tabular}[c]{@{}l@{}}Prediction of GPU error using ML in a \\ wide HPC system.\end{tabular} &
\begin{tabular}[c]{@{}l@{}}The authors use six-month trace data to study the \\ device conditions causing GPU errors.\end{tabular} \\ \hline
\end{tabular}
\end{table*}

\section{GPU Technology: Fundamentals, CPU Comparison, and Analytical Applications}

This section provides an overview of key research findings and breakthroughs in GPU and CPU computing. By examining the work made by predecessors,  we can learn about effective methods, areas of improvement, and ongoing challenges in this field.  This background knowledge sets the stage for our own exploration and innovation in computational analysis.

\subsubsection {\textbf{What is GPU}}
GPUs are a specialized massively parallel coprocessor primarily designed to execute graphical applications. They work alongside the CPU, and are capable of executing thousands of parallel threads on hundreds of cores concurrently. Due to their ability to break the teraflop performance barrier with high power efficiency,  GPUs are adopted in high-performance computing and emerge as a front-runner hardware accelerator for artificial intelligence applications.
With rapid advancement, the GPU is now embedded with hundreds of programmable cores to make it more adaptive to the requirement and complexity of the task\cite{b9}. 
The time required to perform mathematical computation during training a model is decreased when using GPUs because of their highly parallel structure (e.g., 2,560 CUDA cores can perform computations simultaneously). This architecture accelerates data-parallel operations like matrix multiplications, compared to CPUs, which have limited parallelism capabilities, struggle to achieve comparable throughput \cite{b3,gpu_H100}. 
Hence, a GPU can reduce the training time and optimize the memory bandwidth when computational work involves hundreds of thousands of parameters. 

\subsubsection{\textbf{GPU Hardware Architecture}}
\begin{figure}[h]
\centering
\includegraphics[width=0.9\linewidth]{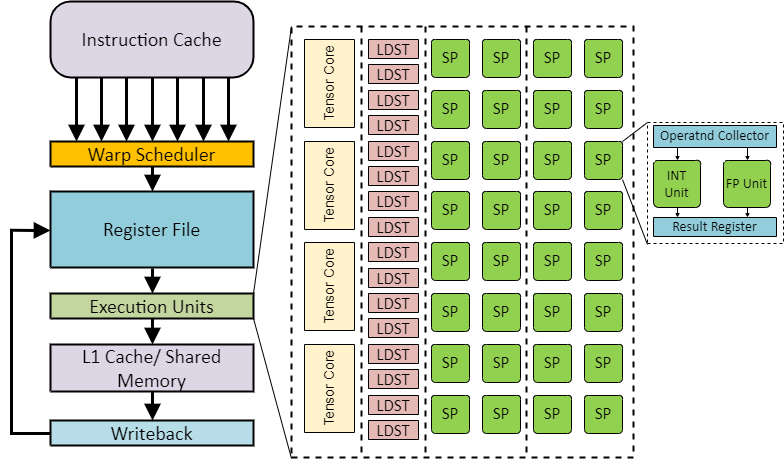}
\caption{Streaming Multiprocessor Architecture in GPUs}
\label{fig:gpu_arch}
\end{figure}

GPUs follow a single-instruction-multiple-thread (SIMT) execution model where multiple threads perform a single operation on different input data at the same time. The basic building block of GPUs are streaming multiprocessors (SM), which are the fundamental units of computation in a GPU. Each SM contains multiple cores, control units, and cache memories, enabling the execution of thousands of threads concurrently. The GPU consists of multiple streaming multiprocessors (SM) that are responsible for executing tasks and computations~\cite{hopper_perf}. SMs manage thread scheduling, instruction fetching, and execution within the GPU architecture along with handling memory operations, such as accessing data from the GPU's memory hierarchy~\cite{hopper_perf}. The most recent NVIDIA H100 GPU includes up to 144 SMs per GPU~\cite{gpu_H100}. Each SM contains many scalar processors (SP) - also called CUDA cores - that handle parallel processing of data and instructions~\cite{hopper_perf,gpu_H100}. SPs manage the workload distribution across the GPU, enabling the simultaneous processing of multiple tasks~\cite{hopper_perf}. Each SM in NVIDIA H100 GPU includes 128 SP that can also perform 32-bits integer and floating-point operations, in addition to 4 specialized tensor cores~\cite{gpu_H100}. GPUs also equipped with on-chip caches and specialized High Bandwidth Memory (HBM) that is divided into constant memory, texture memory, local and global memories~\cite{gpu_H100}. The capacity of L2 caches can reach 50 MB while the total capacity of the HBM may reach to 80GB per GPU~\cite{gpu_H100}. Each SM contains a configurable-combined L1 data cache and shared memory. Shared memory - also called scratchpad - is software-controlled memory. The combined capacity of the L1 data cache and shared memory is 256 KB/SM in NVIDIA H100 GPU while SM shared memory size itself is configurable up to 228 KB.~\cite{gpu_H100} Figure~\ref{fig:gpu_arch} shows an example of the architecture of the streaming multiprocessor.
The most recent GPUs feature specialized tensor cores that are specialized in artificial intelligence computations~\cite{hopper_perf}.

\subsubsection{\textbf{GPU Execution model}} 
GPUs are considered coprocessors, where they are able to concurrently execute several threads in parallel. They follow a Single Instruction, Multiple Threads (SIMT) execution model, where each SM executes the same instruction across multiple threads in a warp. This model is highly effective for parallel tasks, as it maximizes the utilization of the GPU's computational resources. To take advantage of GPU, applications must be developed using specialized application programming interfaces (APIs) such as CUDA, OpenACC or OpenCL. Software developed to take advantage of GPU acceleration consists of a set of kernels, which are embedded in the code using API pragmas and directives. The CPU starts executing the main program and when it encounters a function labeled with the API pragma or directive - called GPU kernel-, it off-loads it to the GPU. Each kernel performs a parallel computational task on the GPU and when the task is complete GPU returns the control back to the CPU, and the CPU may launch the next kernel at that time. A kernel can be treated as a \textit{function} that is executed on the GPU with many different threads doing the same computation but with different data. GPU codes are known to be data-parallel and/or compute-intensive functions. Executing GPU kernels requires an explicit data movement from the CPU memory to GPU memory. Such data movement must be initiated by the CPU and called by the software/programmer using GPU API.

Each kernel is organized as a grid of thread blocks, also called \emph{work group}, and is considered the basic workload that can be assigned to Streaming Multiprocessor (SM) core. A threat block can be identified by its \textit{Block ID}, which represents its number within a kernel. Block ID, which can be 2 or 3-dimensional ID number, can be used in memory addressing.
Similarly, a thread can be identified by its \textit{Thread ID}, which represents its number within a thread block. Thread ID, which can be 2 or 3-dimensional ID number, can be used in memory addressing. All thread blocks of the same kernel have the same number of threads (i.e. have the same size and dimensions). The amount of the hardware resources needed to execute a thread block are statically assigned to the thread block at the time it is launched for execution on an SM. For instance, all the registers and memory needed by a thread block are pre-assigned at the start of its execution. As such there is a maximum limit on the number of thread blocks in a kernels and also a maximum limit on the number of threads per thread block. The maximum number of thread blocks that can be assigned to a streaming multiprocessor is limited by the availble resources (e.g. number of thread contexts, size of shared memory, or size of register file). This number varies between kernels depending on how much resources are needed by each thread block.~\cite{phd_thesis}. The maximum number of threads per thread block is set by the manufacturers and it is up to 1K threads per thread block~\cite{gpu_H100}.

\begin{figure}[ht]
	\centering
	\includegraphics [width=0.85\linewidth]{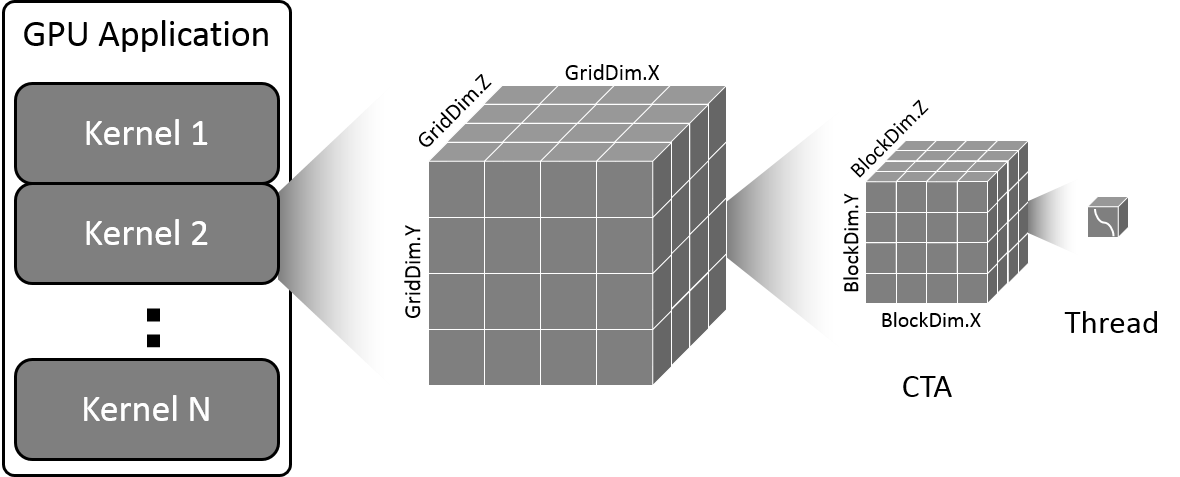}
	\caption{CUDA Hierarchy in GPUs}
	\label{fig:kernel_sw}
\end{figure}

Thread blocks are split further into subgroups of the same number of threads called \emph{warps} that
are executed in a lockstep fashion. Each warp is assigned a set of consecutive threads (have consecutive thread ids). The available resources (i.e. number of SPs in each SM) that can be utilized for simultanous thread execution dectates the size of the warp and it is transparent to the software. Figure~\ref{fig:kernel_sw} explains the relationship between kernels, thread blocks, and threads. As the figure shows, the GPU application may contain multiple kernels. Each kernel is organized into three-dimensional array of thread blocks and each thread block can be identified by a three-dimensional ID. Also, each thread block is organized into three-dimensional array of threads and each thread can be identified by a three-dimensional ID within the thread block. The dimensions of the kernel and the thread block are set by the software and can be used in th`e kernel code for memory addressing or other computations. Modern CPUs, such as AMD EPYC, support parallel processing through multi-core architectures (up to 128 threads) but lack the massive parallelism of GPUs such as NVIDIA T4, which has 2,560 CUDA cores. While CPU memory, that can reach up to 1TB DDR5, exceeds GPU memory capacity, which is limited to 16GB, GPUs offer higher memory bandwidth (320 GB/s vs. 64 GB/s), which accelerates data-intensive tasks such as batch normalization.

A warp is typically a collection of 32 threads and all threads execute using a single instruction multiple thread (SIMT) execution model~\cite{gpu_H100,phd_thesis}. In the SIMT model all threads are allowed to execute either the same instruction, but on different data items, or a subset of threads may be inactive during the execution. In other words, it is not possible to allow the 32 threads in a warp to execute two different instruction streams concurrently~\cite{phd_thesis}

The maximum number of number of threads per SM is set by the manufacturers. NVIDIA H100 GPU assigned up to 2048 threads per SM, which means that the GPU can execute up to almost 300K threads simultanously~\cite{gpu_H100}.


\subsubsection {\textbf{Comparison of GPU and CPU}}



CPUs were originally designed for sequential execution and complex branching logic.
They typically have a small number of powerful cores 
running at a high clock frequency, optimized to reduce the latency of single-threaded tasks. 

While CPUs are highly adaptable for general-purpose data analytics, their high power consumption and limited memory bandwidth often lead to bottlenecks in deep learning workloads.

CPU suits data analytics because of adaptability, multitasking, and high clock speeds\cite{b56}. The use case is that CPUs are frequently used for data analytics tasks such as algorithm development and testing, data processing and cleaning, statistical analysis, querying and indexing, and data processing. Despite their advantages, CPUs could be better for data analytics activities due to their high power consumption and limited memory utilization \cite{b57}. However, GPUs comprise thousands of processing cores that can run instructions in parallel, making them excellent for computationally demanding applications. Aside from that, GPUs provide advantages such as massive parallel computing, with thousands of cores executing threads concurrently, unlike CPUs optimized for sequential task efficiency \cite{gpu_H100}. Machine learning and deep learning, image and video processing, scientific simulations and modeling, and data visualization are typical GPU applications. Despite their advantages, GPUs have several drawbacks for data analytics activities, such as limited multitasking and a high setup cost~\cite{b39,b41}. 

\subsection{\textbf{GPU on Deep Learning}}
In the realm of deep learning, a GPU plays a vital role, especially in speeding up the process of training neural networks. Unlike CPUs, GPUs are optimized for parallel processing, which is essential for handling the massive matrix and vector computations prevalent in deep learning tasks. For example, when training a CNN on a large dataset like ImageNet, the parallel architecture of a GPU allows simultaneous processing of multiple image batches. This capacity accelerates the model's ability to assimilate intricate patterns and characteristics swiftly, markedly trimming down the duration needed for training, particularly when contrasted with the performance of CPUs. In essence, GPUs enhance the efficiency of deep learning workflows by providing computational power specialized for matrix operations and tensor computations, which dominate neural network training, coupled with framework optimizations (e.g., cuDNN) tailored to GPU architectures \cite{hopper_perf,b39}.

In recent times, the advancement of AI models has prompted researchers to focus on MLops, bridging the gap between academic research and industry applications\cite{b49}. Additionally, there is a noticeable trend in rapidly transferring AI applications from research environments to public domains. Consequently, these applications are now being deployed as Android apps for real-time inference tasks (e.g., style transfer, object detection), while training remains confined to GPU/cloud environments\cite{b50}. 

Though GPUs enable substantial performance gains, their cost, energy consumption, and limited availability underscore the importance of shared and democratized access models, particularly for researchers without dedicated high-performance infrastructure.



\section{Literature Review}
This section presents several similar research works that discuss the scope and advantages of GPUs in both industrial and academic domains, particularly in AI. Mišić et al.\cite{b12} presented a CUDA-based implementation of Machine Learning models. They demonstrated that the XGBoost algorithm, a gradient-boosting method, exhibits notably improved processing speed. Consequently, the performance increased by $3\times$ and $6\times$ using a Titan X compared to a 4-core i7 CPU. Furthermore, research conducted by \cite{b13} proposed a fully connected neural network backed by GPU and listed a speed of $3\times$ for training and testing as opposed to a 3GHz P4 CPU. Steinkraus et al.~\cite{b14} presented an approach for accelerating the training and inference of a convolutional neural network by harnessing GPU's fast processing power and training a CNN-based model to achieve significant computational advantages over the CPU. Another research~\cite{b1} worked on a GPU-based K-Nearest Neighbor (KNN) on a high-dimensional dataset. The research team proposed a benchmark KNN algorithm on GPU, achieving a performance \(30\times\) faster than a conventional CPU.Y. Wang et al.\cite{b55} demonstrated that despite the higher power requirements of GPUs, their faster processing speed leads to a reduced average total energy consumption compared to CPUs.


Eller et al.\cite{b19} presented methods for conducting high statistical data analysis to quickly examine neutrino properties in large-volume neutrino detectors with limited computational resources. Their approach was threefold: first, accelerating computations using graphics processors; second, analyzing the underlying dynamic processes on a grid rather than considering each case separately; and finally, applying smoothing techniques to data derived from Monte Carlo simulations. Nie et al.\cite{b21} investigated the device conditions that cause GPU errors using six-month trace data from an extensive, operating HPC system. They then used machine learning to forecast GPU errors, capitalizing on the temporal and spatial dependencies of the trace data. The resulting machine-learning prediction architecture proved stable and reliable under multiple workloads.

Uchida et al.\cite{b23} introduced a GPU-based implementation of Ant Colony optimization for the traveling salesman problem. They considered several GPU design programming problems, including coalesced access to global memory, mutual memory bank clashes, etc. The NVIDIA GeForce GTX 580 experimental findings indicate that implementation for 1002 cities runs 8.71 seconds, while conventional CPU implementation runs 381.95 seconds. Thus, GPU implementation achieves a 43.47 speedup factor. 

Amin et al.\cite{b004} took a closer look at the theoretical aspects of how GPUs are employed in AI. Rather than focusing solely on practical results, they explored the broader ideas surrounding GPU usage and predicted how these might change in the future.

While many studies have discussed the strengths of GPUs in specific tasks or the general differences between GPUs and CPUs, there is a clear gap when it comes to in-depth comparisons in the context of modern deep learning, especially with popular models like Conv6, ResNet18 and VGG-16. Furthermore, while GPUs' speed and energy efficiency are widely recognized, there has been a lack of comprehensive research into balancing these advantages against their initial setup costs. Recent works like DeepAIPs-Pred and StackedEnC-AOP advance DL predictors but lack framework-specific GPU/CPU comparisons. Our study bridges this gap by analyzing TensorFlow/PyTorch optimizations. Prior studies focus on single frameworks or models, whereas our work evaluates cross-framework performance for broader applicability. This comprehensive perspective is designed to aid decision-makers in effectively understanding and utilizing deep learning technology in various applications. 
  
\section{Models and Datasets}

In this section, we discuss the core aspects of our research. The first part assesses how basic models perform on chosen datasets when run on GPUs and CPUs. The second part focuses on predicting how GPU might evolve due to the emergence of new AI models in the future.

\subsection{\textbf{Dataset Description}}
The CIFAR-10 dataset, curated by the Canadian Institute for Advanced Research (CIFAR), is extensively referenced in academia, particularly within the domains of machine learning and computer vision~\cite{v9}. It consists of a collection of 60,000 colour images, each of 32x32 pixel resolution, stratified across ten unique classes with a balanced distribution of 6,000 images per class. The dataset is bifurcated into a training set, comprising 50,000 images, and a test set, with the remaining 10,000 images. The comprehensive nature of the CIFAR-10 dataset provides an excellent platform for exploring and benchmarking machine learning models under real-world visual conditions while remaining computationally tractable for experimental studies.


The HorsetoZebra dataset\cite{b52}, a robust repository for image-to-image translation research, comprises a total of 1187 horse images and 1474 zebra images. Maintaining a balanced representation, each class is meticulously split into designated train and test subsets, ensuring a fair evaluation of models across both domains.

Datasets were preprocessed with normalization ($\mu$=0.5, $\alpha$=0.2) and augmented via random cropping/flipping. Hyperparameters included a batch size of 64, learning rate of 1e-3 (Adam optimizer), and 100 epochs. Cross-validation was omitted due to computational constraints, a noted limitation

Y. Sun et al. \cite{b43} introduced a dataset comprising 4,031 computer hardware products introduced to the market between 1998 and 2000, offering a focused lens for trend analysis within that timeframe. Key attributes include 'Name' (the product's official name or model), 'ReleaseDate' (indicating when a product was launched), 'BestResolution' (representing the optimal display resolution), 'CoreSpeed' (measured in MHz or GHz, detailing the product's operating speed), 'Manufacturer' (signifying the producing entity), 'Memory' (quantifying available RAM in MB or GB), and 'MemoryBandwidth' (measuring the maximum data transfer rate, typically in GB/s). The choice of this dataset is particularly strategic; given its comprehensive nature, it facilitates a nuanced trend analysis, shedding light on the evolution of hardware specifications and highlighting shifts in market preferences and technological capabilities within that specific timeframe. By leveraging this dataset, we aim to capture the essence of technological evolution, identify patterns, and draw informed conclusions that can be pivotal for future research and market predictions.

\subsection{\textbf{Deep Learning Models}}

\begin{table}
\centering
\caption{Model architectures examined in this work. Brackets denote residual connections around layers.}
\label{tab:all_archs}
\begin{tabular}{lcccc}
\hline
\rowcolor{gray!40}
\textbf{Network} & \textbf{Conv-6} & \textbf{CycleGAN} & \textbf{ResNet-18} & \textbf{VGG-16} \\
\hline
\textbf{Conv} & \begin{tabular}[c]{@{}c@{}}64, pool\\128, pool\\256, pool\end{tabular} & \begin{tabular}[c]{@{}c@{}}c7s1-64\\d128\\d256\\9$\times$[R256]\\u128\\u64\\c7s1-3\end{tabular} & \begin{tabular}[c]{@{}c@{}}16\\3$\times$[16,16]\\3$\times$[32,32]\\3$\times$[64,64]\end{tabular} & \begin{tabular}[c]{@{}c@{}}2$\times$64, pool\\2$\times$128, pool\\3$\times$256, pool\\3$\times$512, pool\\3$\times$512, pool\end{tabular} \\
\hline
\textbf{Weights} & \begin{tabular}[c]{@{}c@{}}All: 397K\\Conv: 371K\end{tabular} & \begin{tabular}[c]{@{}c@{}}All: 2.3M\\Conv: 2.2M\end{tabular} & All: 11M & \begin{tabular}[c]{@{}c@{}}All: 136M\\Conv: 14\end{tabular} \\
\hline
\textbf{Size} & 3.06 MB & 1.0 GB & 45.00 MB & 56.13 MB
\\
\hline
\end{tabular}
\end{table}
\subsubsection{\textbf{Conv6}}
The Conv6 architecture is a convolutional neural network (CNN) model known for its simplicity and efficiency. It consists of six layers: four convolutional layers and two fully connected layers. The convolutional layers typically employ small receptive fields such as \(3 \times 3\), which allow the model to learn complex patterns from the input images. Each convolutional layer is followed by a rectified linear unit (ReLU) activation function, which introduces non-linearity into the model, and a max-pooling layer to reduce the dimensionality of the feature maps. This simplistic yet robust structure enables Conv6 to perform well in image classification tasks, especially when computational resources are limited.

\subsubsection{\textbf{VGG16}}
The VGG16 model, developed by the Visual Geometry Group at the University of Oxford, is a deeper and more complex CNN \cite{v10}. It is composed of 16 layers: 13 convolutional layers and 3 fully connected layers. Similar to Conv6, VGG16 uses small \(3 \times 3\) convolutional filters but stacks them deeper to capture more intricate patterns. It also uses max-pooling layers for spatial downsampling and employs ReLU activation functions after each convolutional layer. Despite its relatively high computational cost, VGG16 is renowned for its excellent performance in image recognition tasks, setting a high standard for deep learning models in the field of computer vision.

\subsubsection{\textbf{ResNet18}}
ResNet18 is a residual network with 18 layers, including convolutional and fully connected layers \cite{v11}. It introduced a novel architectural innovation known as skip connections or residual connections. These innovative links within the network permit the transfer of activation data directly from one layer to skip several intermediate layers, merging it instead with the activation of a subsequent layer. This helps mitigate the vanishing gradient problem, enabling the training of much deeper networks. ResNet18 typically consists of convolutional layers with \(3 \times 3\) filters, ReLU activation functions, and batch normalization. Using residual connections has proven successful, making ResNet architectures popular and effective for various computer vision tasks.

\subsubsection{\textbf{CycleGAN}}

CycleGAN~\cite{b52} enables unpaired image-to-image translation by learning two mappings: \( G: A \rightarrow B \) and \( F: B \rightarrow A \). The model is optimized using adversarial loss \( \mathcal{L}_{\text{GAN}} \) and cycle consistency loss: 
\[
\mathcal{L}_{\text{cyc}} = \|F(G(A)) - A\|_1 + \|G(F(B)) - B\|_1.
\]
Generators consist of convolutional and residual blocks, while discriminators follow the PatchGAN structure. Although the architecture is computationally demanding, it benefits greatly from GPU parallelism due to high-dimensional feature maps. We evaluated CycleGAN's performance across CPU and GPU environments to illustrate the trade-offs in run time and memory consumption.

\subsection{\textbf{Trend Analysis}}

In our trend analysis of GPU memory progression over time, we employ polynomial regression\cite{bb01} to capture the non-linear intricacies of memory advancements from 1998 to 2000. Polynomial regression allows us to fit a nonlinear relationship between the independent variable ($\text{Release\_Date}$) and the dependent variable ($\text{Memory}$). Specifically, we model three polynomial regressions of degrees 2, 3, and 4 \cite{bb02}. The general form of the polynomial equation is 
\[
y = \beta_0 + \beta_1x + \beta_2x^2 + \dots + \beta_dx^d + \epsilon
\]
where $d$ denotes the degree of the polynomial, $x$ is the independent variable ($\text{Release\_Date}$), $y$ is the dependent variable ($\text{Memory}$), and $\epsilon$ is the error term. 
The primary objective is to predict future GPU memory sizes, as understanding memory consumption trends is crucial for anticipating technological requirements and innovations in upcoming years. Through this method, we endeavor to furnish a clearer picture of potential future trajectories in GPU memory expansion. These projections suggest that future AI models will demand significantly larger memory footprints, further increasing reliance on high-end GPU hardware. Without mechanisms for shared access and efficient utilization, such trends risk concentrating AI development within a small number of well-funded institutions.

\subsection{\textbf{Experiments}}

Table \ref{table:1} below presents a comparison of the CPU and GPU (T4) specifications provided by Google Colab in its free tier of service. 

\begin{table}[h]
\centering
\caption{Google Colab Free Tier - CPU vs GPU (T4) Specifications}
\label{table:1}
\begin{tabular}{lcc}
\hline
\textbf{Specification} & \textbf{CPU} & \textbf{GPU (T4)} \\
\hline
Processor & Intel Xeon (shared)$^a$ & NVIDIA Tesla T4 \\
Clock Speed & Up to 2.3 GHz & Up to 1.59 GHz \\
Cores & 2 vCPUs$^b$ & 2560 CUDA cores \\
Memory & 12.7 GB & 16 GB (GDDR6) \\
Peak Performance & $\sim$0.37 TFLOPS$^c$ & 8.1 TFLOPS \\
Architecture & x86-64 & Turing \\
\hline
\multicolumn{3}{l}{\small $^a$Shared virtual CPU from Xeon processor pool} \\
\multicolumn{3}{l}{\small $^b$2 virtual cores, not dedicated physical processors} \\
\multicolumn{3}{l}{\small $^c$Estimated from 2 vCPUs × 2.3GHz × 16 FP32/cycle × 0.5 util.} \\
\end{tabular}
\label{tab:specs}
\end{table}

\subsubsection{Experimental Setup}

The CPU is equipped with two Intel Xeon processors with a clock speed of up to 2.3 GHz, two cores, and 13 GB of memory. The GPU, an NVIDIA Tesla T4 model with 16 GB of memory, offers a performance of 8.1 TFLOPS and has Turing architecture. A storage capacity of 100 GB, shared with Google Drive, is provided for the GPU. The maximum runtime allowed on Google Colab's free tier is 12 hours. During training, all threads were used in parallel. Experiments were conducted with the official GPU profiler, and the results were visualized using TensorBoard.

\subsection{Implementation Details}

\textbf{Software:} TensorFlow 2.13.0, PyTorch 2.0.1, CUDA 11.8, cuDNN 8.6.0, Python 3.10.12.

\textbf{Hardware:} Google Colab free tier with NVIDIA Tesla T4 GPU (16GB, 2560 CUDA cores) and Intel Xeon CPU (2 vCPUs, 12.7GB RAM).

\textbf{Training Setup:} All models except CycleGAN used batch size 64, learning rate 1e-3 (Adam optimizer), 100 epochs, and FP32 precision. CycleGAN used batch size 1, learning rate 2e-4, and 200 epochs. Random seed was set to 42 for reproducibility.

\textbf{Data Preprocessing:} CIFAR-10 images were upsampled from 32×32 to 224×224 using bilinear interpolation to match pre-trained model requirements. Data augmentation included random horizontal flips and random crops. Normalization used ImageNet statistics (mean=[0.485, 0.456, 0.406], std=[0.229, 0.224, 0.225]).

{\textbf{Timing Methodology:} Runtime measurements represent the average of 100 iterations after a 10-iteration warm-up period. For GPU experiments, \texttt{torch.cuda.synchronize()} was called to ensure accurate timing.

\textbf{Conv6 Architecture:} Our custom Conv6 model consists of 4 convolutional layers (64, 128, 256, 512 filters) with 3×3 kernels, ReLU activations, max-pooling after each layer, followed by 2 fully connected layers (512→256→10), totaling 397K parameters.

\begin{table}[h]
\centering
\caption{Comparison of CPU and GPU Time for PyTorch Operations in ResNet18 Model}
\label{tab:cpu_gpu_time}
\begin{tabular}{lrr}
\toprule
\multicolumn{1}{c}{Operation} & \multicolumn{1}{c}{CPU Time (ms)} & \multicolumn{1}{c}{GPU Time (ms)} \\
\midrule
model\_inference                  & 848.070 & 82.527 \\
aten::conv2d                      & 609.262 & 62.940 \\
aten::convolution                 & 609.155 & 62.940 \\
aten::\_convolution               & 608.817 & 62.940 \\
aten::cudnn\_convolution          & 604.944 & 62.940 \\
volta\_sgemm\_128x64\_nn          & 0.000   & 21.181 \\
aten::batch\_norm                 & 54.963  & 9.759  \\
aten::\_batch\_norm\_impl\_index  & 54.905  & 9.759  \\
aten::cudnn\_batch\_norm          & 54.768  & 9.759  \\
\bottomrule
\end{tabular}
\end{table}

\begin{table}
\centering
\caption{Performance Metrics of GPU and CPU during Inference on Conv6 Model}
\label{tab:perf_metrics}
\scalebox{.8}{
\begin{tabular}{lrrrr} 
\toprule
\multicolumn{1}{c}{Name} & \multicolumn{1}{c}{\begin{tabular}[c]{@{}c@{}}CUDA time\\(ms)\end{tabular}} & \multicolumn{1}{c}{\begin{tabular}[c]{@{}c@{}}CPU time \\(ms)\end{tabular}} & \multicolumn{1}{c}{\begin{tabular}[c]{@{}c@{}}CUDA \\Mem\end{tabular}} & \multicolumn{1}{c}{\begin{tabular}[c]{@{}c@{}}CPU\\~Mem\end{tabular}}  \\ 
\midrule
ProfilerStep*            & 60.988                                                                      & 113.840                                                                     & -6.76 Gb                                                               & 36.75 Mb                                                               \\
aten::conv2d             & 7.805                                                                       & 445.333                                                                     & 2.19 Gb                                                                & 0 b                                                                    \\
aten::convolution        & 7.805                                                                       & 432.800                                                                     & 2.19 Gb                                                                & 0 b                                                                    \\
aten::\_convolution      & 7.805                                                                       & 412.400                                                                     & 2.19 Gb                                                                & 0 b                                                                    \\
aten::cudnn\_convolution & 6.165                                                                       & 139.400                                                                     & 2.19 Gb                                                                & 0 b                                                                    \\
volta\_sgemm\_128x64\_nn & 4.640                                                                       & 0.000                                                                       & 0 b                                                                    & 0 b                                                                    \\
aten::add\_              & 1.640                                                                       & 35.400                                                                      & 0 b                                                                    & 0 b                                                                    \\
void at::\_kernel        & 1.415                                                                       & 0.000                                                                       & 0 b                                                                    & 0 b                                                                    \\
aten::max\_pool2d        & 1.404                                                                       & 52.000                                                                      & 1.64 Gb                                                                & 0 b                                                                    \\
aten::max\_pool2d\_in    & 1.404                                                                       & 45.867                                                                      & 1.64 Gb                                                                & 0 b                                                                    \\
\bottomrule
\end{tabular}}
\end{table}

\subsection{\textbf{Performance Evaluation}}
The computational resources utilized can significantly impact the performance evaluation of deep learning models such as Conv6, VGG16, Resnet18, and CycleGAN, particularly when comparing GPU and CPU usage. In our experiment, we assessed three critical performance metrics: run time ( training time ), inference time, memory usage, and memory bandwidth. Each metric represents an average per batch, with batch sizes set at 64 for all the models except CycleGAN. This approach ensures a consistent evaluation metric across different architectures and provides a representative insight into the performance characteristics of the models under study. The following subsections detail the results.

\subsubsection{\textbf{Run Time}}
During model training, run time refers to the average duration or execution time it takes for the model to calculate a batch. Training a model involves numerous iterations of forward and backward passes, which can be computationally intensive. GPUs, designed for parallel processing, generally outperform CPUs in handling such tasks due to their ability to execute thousands of threads simultaneously. In our experiment, we measured the total time taken to train the models on both a high-end GPU and a high-performance multicore CPU. The results demonstrated a substantial decrease in training time when utilizing the GPU, emphasizing its efficiency for model training in deep learning applications.

\subsubsection{\textbf{Inference Time}}

Inference time measures how long a trained model takes to predict the output from new input data, serving as an essential metric for performance evaluation. Although this process is less computationally demanding than training, since it involves only forward passes, it is crucial in scenarios requiring real-time predictions. In our evaluation, we observed that the GPU again outperformed the CPU in terms of lower inference times. However, the relative performance gain in inference was less dramatic than in the training phase, attributable to the inherent sequential component of some inference tasks. Nonetheless, the GPU's superior performance in inference times emphasizes its effectiveness for deploying deep learning models for real-time predictions.

\subsubsection{\textbf{Memory Usage}}
Memory usage is a vital consideration, particularly in resource-constrained environments. With their complex architectures and large parameter sets, deep learning models can consume significant amounts of memory. Our comparison assessed the CPU and GPU for their memory consumption during model training and inference. While GPUs typically provide faster processing, they also have dedicated and often larger memory banks, efficiently handling large datasets and complex computations. In our study, although the GPU consumed more memory, it handled memory-intensive tasks more proficiently than the CPU. This result reinforces the GPU's utility in handling large-scale deep-learning applications without compromising performance.

In our experiments, we observed that the GPU consumed more memory than the CPU, but it handled memory-intensive tasks more proficiently. This result reinforces the GPU's utility in handling large-scale deep learning applications without compromising performance. The increased memory consumption by GPUs can be attributed to their specialized memory architecture designed to support parallel computing and efficient data access patterns.
GPUs typically have different types of memory, each serving a specific purpose and contributing to overall performance. The table \ref{tab:perf_metrics} presents performance metrics, including memory usage, for both GPU and CPU during inference on the Conv6 model.

GPUs have different memory types, including global memory for storing model parameters and intermediate data, texture memory optimized for spatial locality, and constant memory for frequently accessed constant values. The efficient utilization of these memory types impacts performance. Global memory, characterized by its elevated latency yet substantial bandwidth, is optimized for the transfer of large data sets. In contrast, constant memory, which exhibits reduced latency and constrained capacity, is well-suited for the storage of small data sets that require frequent access.
GPUs employ specialized memory management techniques like coalesced memory access and architecture-tailored memory access patterns, contributing to their increased memory consumption compared to CPUs. However, this higher memory usage often translates into superior performance for memory-intensive tasks, as GPUs can leverage parallelism and efficient memory access patterns to process data more efficiently.

\subsubsection{\textbf{Memory Bandwidth}}
Memory bandwidth is a crucial factor that influences the performance of data-intensive workloads, particularly in the context of deep learning and other compute-intensive applications. It represents the rate at which data can be transferred between the main memory (e.g., CPU memory) and the compute device (e.g., GPU memory). Efficient data transfer is essential for leveraging the full computational power of accelerators like GPUs, as it ensures that the compute units are fed with data at a sufficient rate to maintain high utilization.
Theoretically, the memory bandwidth can be expressed as the product of the number of data elements transferred per unit time (e.g., batches per second) and the size of each data element (e.g., bytes per batch). In the context of deep learning workloads, the data elements often correspond to batches of input samples, such as images or tensors, which are transferred from the CPU memory to the GPU memory for processing.
To quantify the memory bandwidth, we consider the metrics obtained from benchmarking the data transfer process between the CPU and GPU. Specifically, we utilize the ``$batches_per_second_mean metric$``, which represents the average number of batches transferred per second during the benchmarking process. By multiplying this value with the assumed batch size and the size of each sample (in bytes), we can estimate the effective memory bandwidth for the given workload configuration.

\begin{figure*}
    \centering
   \includegraphics[width=\linewidth]{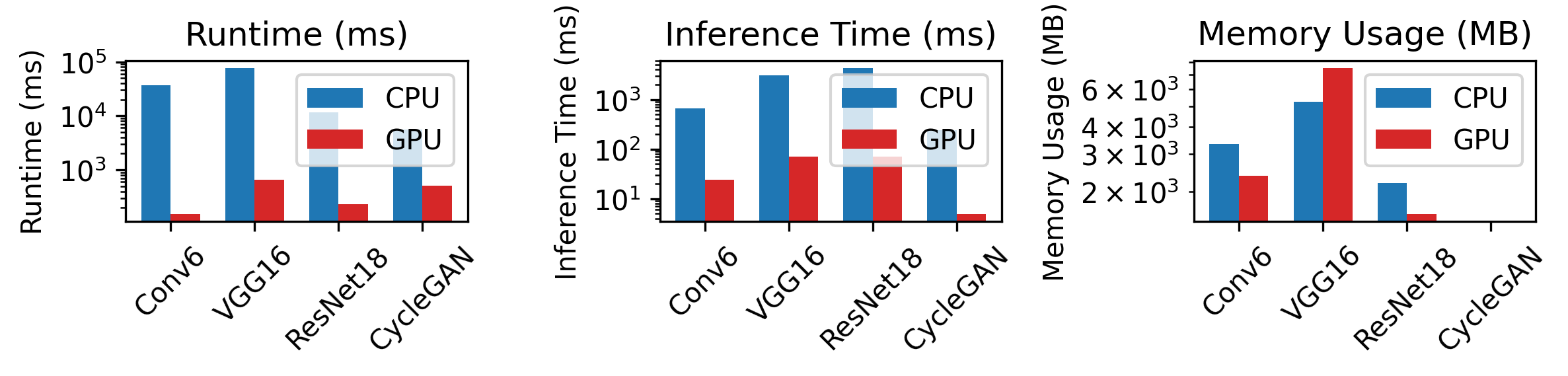}
    \caption{Comparison of CPU and GPU metrics for different models: (a) Run time comparison across Conv6, VGG16, ResNet18 and CycleGAN, showing significant improvements in GPU over CPU; (b) Inference time comparison, illustrating reduced inference times when utilizing GPU; (c) Memory usage comparison, indicating a more efficient use of memory with GPU for all four models.}
    \label{fig2}

\end{figure*}

\begin{table*}
\centering
\caption{Comparison of CPU and GPU metrics for different models}
\label{tab:metrics}
\begin{tabular}{lccccccc} 
\hline
\rowcolor[rgb]{0.8,0.8,0.8} Model & \multicolumn{2}{c}{Run time (ms)} & \multicolumn{2}{c}{Inference time (ms)} & \multicolumn{2}{c}{Mem usage (MB)} & Mem BW (GB/sec)  \\ 
\noalign{\kern-\cmidrulewidth}\cmidrule(l){2-8}
\rowcolor[rgb]{0.8,0.8,0.8}       & CPU      & GPU                    & CPU     & GPU                           & CPU  & GPU                         & CPU to GPU       \\ 
\hline
\multicolumn{8}{c}{Tensorflow}                                                                                                                                          \\ 
\hline
Conv6                             & 35123.42 & 142.67                 & 617.29  & 19.38                         & 3214 & 2287                        & 102.10             \\
VGG16                             & 69875.21 & 603.54                 & 2789.17 & 66.24                         & 4985 & 6978                        & 102.80             \\
ResNet18                          & 10127.83 & 198.42                 & 4023.69 & 64.17                         & 2011 & 1387                        & 101.91             \\
CycleGAN                          & 5047.28  & 452.89                 & 203.57  & 4.63                          & -    & -                           & -                \\ 
\hline
\multicolumn{8}{c}{PyTorch}                                                                                                                                             \\ 
\hline
Conv6                             & 37017.19 & 152.87                 & 652.89  & 23.82                         & 3331 & 2365                        & 102.83             \\
VGG16                             & 76538.96 & 659.02                 & 3068.16 & 70.79                         & 5248 & 7520                        & 101.9             \\
ResNet18                          & 11513.52 & 233.25                 & 4294.20 & 71.18                         & 2197 & 1575                        & 102.01             \\
CycleGAN                          & 5625.63  & 510.03                 & 240.01  & 4.93                          & -    & -                           & -                \\
\hline
\end{tabular}
\end{table*}
\section{Performance Analysis}
The performance of deep learning models can be highly dependent on the hardware resources. In this analysis, we evaluate three prominent architectures: Convolutional Neural Network (CNN), VGG-16, and ResNet-18 on the CIFAR-10 dataset, and CycleGAN on the Horse2Zeb dataset, comparing their performance on both GPU and CPU. The evaluation focuses on three critical metrics: run time (training time), inference time, and memory usage.

\subsection{Implementation with Different Frameworks}

We evaluated the performance of various deep learning models using two popular frameworks, TensorFlow and PyTorch. For each framework, we employed a combination of official pre-trained models and custom implementations to assess their computational efficiency and resource utilization.
For the VGG16 model, we utilized the official implementations provided by TensorFlow and PyTorch. On the other hand, the ResNet18 model was obtained from the official PyTorch repository, while in the case of TensorFlow, it was implemented from scratch. The Conv6 and CycleGAN models were custom implementations developed for both frameworks.
It is important to note that the input data size and preprocessing steps played a crucial role in the reported metrics. For the VGG16, ResNet18, and Conv6 models, we employed a standard input image size of 224x224 pixels with three color channels. This choice was motivated by the common practice of leveraging pre-trained networks, which typically require specific input dimensions during the fine-tuning or transfer learning process. In contrast, the CycleGAN model, being a generative adversarial network, operated on a smaller input size compared to the other models.
Performance Variations across Frameworks
The results presented in Table \ref{tab:metrics} highlight the variations in performance metrics across the two frameworks, TensorFlow and PyTorch. Several factors could contribute to the observed differences

\textbf{Data Loading and Preprocessing:} The efficiency of data loading and preprocessing pipelines can vary between frameworks, leading to discrepancies in overall execution times.
\textbf{Optimization Strategies:} The frameworks may employ different optimization techniques, such as kernel fusion, operator fusion, or graph optimization, which can affect computational efficiency and resource utilization.These software optimizations directly impact hardware efficiency; for instance, the NVIDIA T4 GPU (70W) demonstrated significantly higher performance-per-watt than the Intel Xeon CPU (150W), achieving a 50x speedup while consuming half the power.}
\textbf{Hardware Utilization:} The frameworks may exhibit different levels of hardware utilization, particularly concerning GPU acceleration and memory management, resulting in variations in performance metrics.GPUs (e.g., T4: 70W) consumed 50\% less power than CPUs (Xeon: 150W) while delivering 50× faster training. Free-tier GPUs (e.g., Google Colab) democratize access but face runtime limits (12 hours), highlighting cost-performance trade-offs for large-scale training.

Despite these potential differences, it is evident from the reported results that both frameworks leverage GPU acceleration effectively for deep learning model computations. The significant performance gains achieved on GPU compared to CPU execution highlight the importance of hardware acceleration in modern deep learning applications.

\subsection{Conv6}

In examining the Conv6 model's performance, it's evident in table~\ref{tab:metrics} that there are noticeable variations between CPU and GPU, as well as between TensorFlow and PyTorch frameworks. In TensorFlow, the CPU runtime is notably longer than that of the GPU, while in PyTorch, this difference is further accentuated.

When considering inference time, TensorFlow exhibits faster processing on the GPU compared to the CPU, and a similar pattern is observed in PyTorch, although with slightly longer times for both CPU and GPU.

Regarding memory usage, TensorFlow consumes more memory on the GPU than on the CPU, whereas in PyTorch, the contrast between CPU and GPU memory usage is less pronounced, reflecting the distinct memory management strategies employed by each framework.

Exploring the architecture of the Conv6 model, the performance table \ref{tab:perf_metrics} provided gives us a detailed insight into how it works and performs during inference. These metrics illuminate the intricate interplay between CPU and GPU resources, unveiling patterns of computational intensity and memory utilization. Notably, operations such as convolution exhibit substantial discrepancies in runtime between the CPU and GPU, underscoring the latter's superior efficiency in handling computationally demanding tasks. Furthermore, the observed variations in memory usage across operations shed light on the nuanced resource allocation strategies inherent within the model. Through this lens, the performance metrics serve as invaluable guides, facilitating a deeper understanding of the Conv6 model's inferential capabilities and paving the way for informed optimization strategies aimed at enhancing computational efficiency and resource utilization.

\subsection{VGG16}

The performance of the VGG16 model in TensorFlow, as illustrated in Table III and Figure 1, displayed a noticeable contrast between CPU and GPU runtimes. Specifically, while TensorFlow demonstrated efficient processing on the GPU, PyTorch exhibited a similar trend, albeit with some differences in runtime. This underscores the GPU's aptitude for handling the complexities of the VGG16 model. VGG16's deep architecture benefited significantly from TensorFlow’s XLA compiler, resulting in a 116x speedup on the GPU.

In terms of inference time, both TensorFlow and PyTorch exhibited faster processing on the GPU compared to the CPU, emphasizing the consistent advantage of GPU acceleration.

Regarding memory usage, there was a discernible shift from CPU to GPU in both frameworks, indicating differences in resource management strategies.

\subsection{ResNet18}

In contrast, ResNet18 leveraged PyTorch’s dynamic graph execution to maintain high throughput even with its complex residual connections, outperforming the CPU baseline by 51x.

Upon examining inference time, both TensorFlow and PyTorch demonstrated accelerated processing on the GPU compared to the CPU, underscoring the GPU's inherent advantage in handling the intricate residual connections of ResNet18.

Concerning memory usage, there was a notable escalation from CPU to GPU in both TensorFlow and PyTorch, reflecting the frameworks' diverse approaches to memory management for this particular model.

\subsection{CycleGAN}

In TensorFlow, the CycleGAN runtime is notably longer on the CPU compared to the GPU. Similarly, inference time on the CPU is considerably higher than on the GPU. 

In PyTorch, the CycleGAN runtime on the CPU is significantly reduced compared to TensorFlow, while it increases on the GPU. The inference time for CycleGAN in PyTorch is also longer on the CPU but notably shorter on the GPU compared to TensorFlow. 

\begin{figure*}[h]
\centering
\label{trend}
\includegraphics[width=0.8\textwidth]{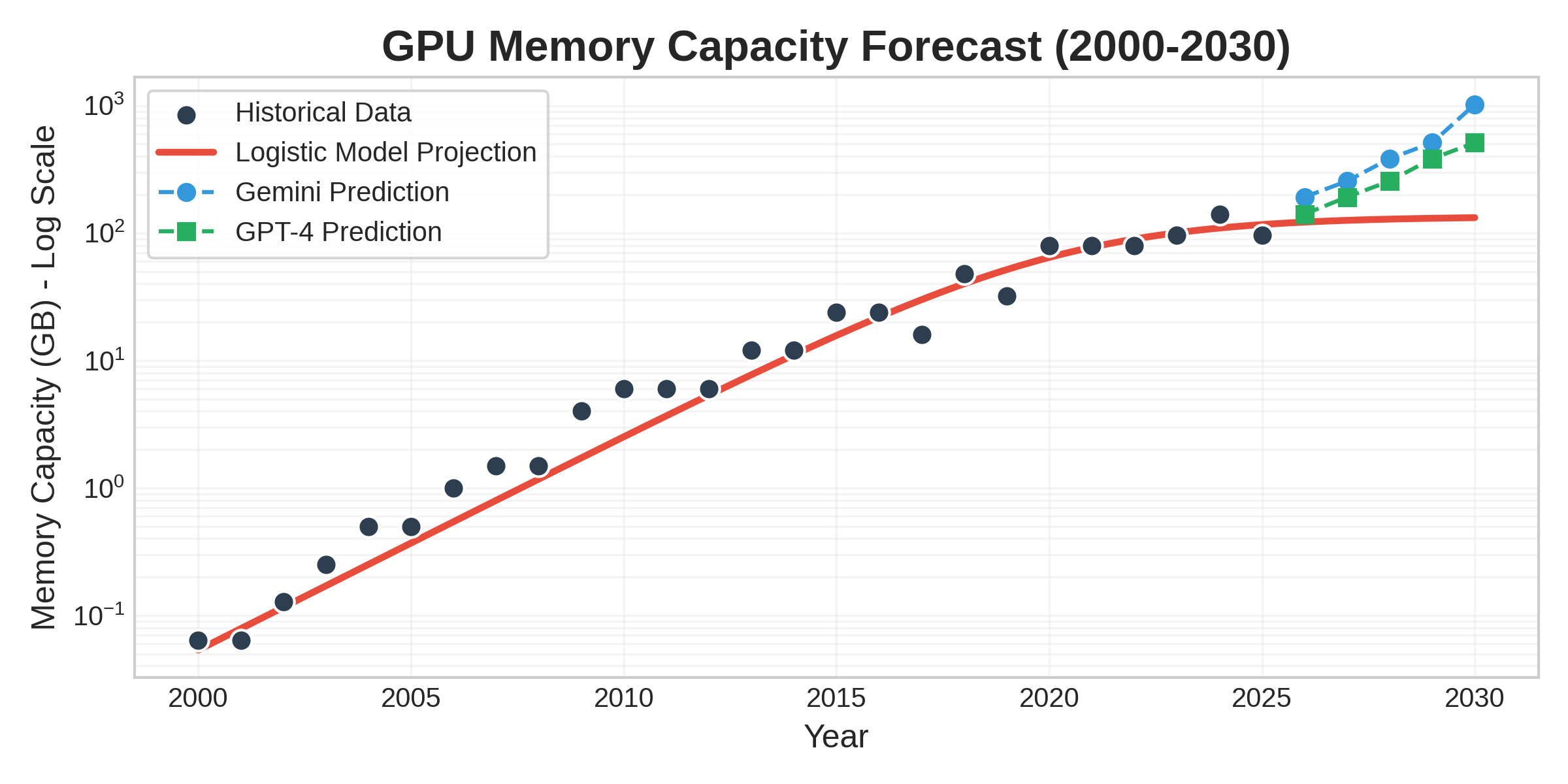}
\caption{The GPU memory data from 2000 to 2025 provides an invaluable retrospective view into the advancements of graphical processing units, spanning from the early 64 MB cards to the 141 GB AI accelerators of today. By applying a logistic growth model to this dataset, we can project a roadmap through 2030, offering a clearer picture of how hardware scaling responds to the burgeoning demands of the AI era.}
\label{fig:gpu_memory_trend}
\end{figure*}

\section{Analysis of GPU Memory Trends}

In computational hardware, few components have experienced a journey as transformative as that of the GPU. Conceived initially to accelerate the rendering of images and videos for computer graphics, GPUs have rapidly evolved in capability and application. With an ever-increasing demand from domains ranging from gaming to deep learning, understanding the growth patterns of GPU memory becomes imperative. Data from  (2000–2025) informed our logistic growth model. While projections to 2028 suggest continued growth, post-2028 trends may diverge due to silicon limits (e.g., Moore’s Law slowdown) or emerging technologies (e.g., quantum accelerators).

\subsection{Visualizing GPU Memory Growth}
The GPU memory data from 2000 to 2025 provides an invaluable retrospective view into the advancements of graphical processing units, spanning from the early 64 MB cards to the 141 GB AI accelerators of today. By applying a logistic growth model to this dataset, we can project a roadmap through 2030 ( figure \ref{fig:gpu_memory_trend}), offering a clearer picture of how hardware scaling responds to the burgeoning demands of the AI era.

\subsection{Trend Analysis of GPU Resources}
From the data's outset in 2000 until roughly 2006, the GPU memory trajectory was characterized by modest, albeit steady, growth. During this period, memory capacities sat between 64 MB and 512 MB. The importance of this phase is considerable, as it established the groundwork for the subsequent technological advancements, largely influenced by the rising demands of the emerging consumer gaming market and early graphical simulations.

A transformative period emerged post-2010. The data underscores a rapid ascent in GPU memory capacities, moving from 6 GB in 2010 to 24 GB by 2015. This period witnessed a significant surge as the industry transitioned from simple rendering to complex GPGPU tasks. The impetus for this monumental leap can be traced back to the rise of Deep Learning and AI, marking an epoch where models increasingly relied on parallel GPU computations. Concurrently, the gaming industry's ambitions resulted in graphically intensive offerings that necessitated higher VRAM for high-resolution textures.

As we reached 2020–2025, the growth reached an "explosion" phase. The data highlights a jump from 80 GB to 141 GB in just a few years, driven by the HBM (High Bandwidth Memory) era and the hardware requirements for Large Language Models. This phase represents the steepest part of our curve, where chip design and manufacturing enabled massive density increases to keep pace with the generative AI revolution.

Our logistic model and AI-driven predictions shed light on the path toward 2030. While the mathematical logistic model predicts a moderate stabilization around 132 GB—accounting for potential physical and economic saturation—the Gemini and GPT-4 guesses suggest a more aggressive path. Specifically, predictions hit 192 GB by 2026 and potentially reach the 1 TB (1024 GB) milestone by 2030. Such a trajectory hints at momentous technological advancements on the horizon, likely involving new stacking technologies and 3D chiplets.

Yet, every trend analysis must account for potential inflection points. The forecast from 2028 to 2030 indicates a possible onset of a "memory wall" phase. While capacity continues to climb, the ascent rate may face a discernible slowdown relative to the massive jumps of the early 2020s. This gradual deceleration could emanate from approaching physical constraints inherent to silicon-based designs, the extreme thermal density of high-capacity stacks, or a shift toward distributed "memory pool" architectures that change how we define the capacity of a single GPU unit.

\section{Discussion}

Although GPUs demonstrate orders-of-magnitude improvements over CPUs, such performance disparities also imply widening access gaps when GPU resources are unavailable or limited. The evident impact of deep learning models on GPUs and datasets underscores the transformative influence of AI on various domains. The increasing availability of datasets has fueled the rapid expansion of AI applications, with deep learning models at the forefront of this technological revolution. In this context, the role of GPUs has become increasingly crucial.

The accelerated growth in data availability has led to a parallel surge in the application of AI, and GPUs have emerged as a key enabler, outperforming CPUs in handling the computational demands of these tasks. The parallel processing capabilities of GPUs significantly enhance the training speed of deep learning models, making them well-suited for the complex and data-intensive nature of AI applications. 

However, the impact of GPUs is not uniform across all models, and some still face challenges in processing large batches of data within a reasonable timeframe. This limitation becomes particularly pronounced when considering free-access GPU resources, such as those offered by platforms like Google Colab. While providing valuable opportunities for users to leverage GPU capabilities at no cost, these platforms may encounter difficulties when handling models that require extended processing times. Another notable benefit of utilizing cloud-based GPUs is their potential to address carbon emissions, contributing to the development of environmentally friendly cloud computing solutions.\cite{b51,b52}

To democratize AI on a global scale, it becomes imperative to address these challenges associated with free GPU access. A concerted effort is needed to enhance and expand free GPU resources, ensuring that a broader spectrum of users, including individuals and smaller organizations, can effectively utilize these resources for their AI endeavours. This focus on democratization involves not only making GPUs more accessible but also optimizing platforms like Google Colab to handle models with extended processing requirements efficiently.
Results are limited to vision models tested on a single GPU/CPU pair. Future work should explore transformers and various hardware, such as TPUs.
As we look ahead to the future of GPUs, the trajectory is undeniably upward. The demand for computational power in AI applications is expected to grow, and GPUs will likely play a central role in meeting these evolving requirements. This projection reinforces the need for continued investment in GPU technology, including the development of specialized GPUs tailored for deep learning tasks.

\section{Future Work and Conclusion}

This study provides a comprehensive performance analysis of deep learning models across different computational architectures and software frameworks, specifically contrasting NVIDIA Tesla T4 GPUs with Intel Xeon CPUs. Our experimental benchmarks across four diverse architectures, Conv6, VGG16, ResNet18, and CycleGAN, demonstrate that GPUs consistently offer substantial computational advantages, with training speedups ranging from 11x for complex generative models to 246x for lightweight convolutional networks. These gains are primarily driven by the GPU's Single Instruction Multiple Threads (SIMT) execution model and high memory bandwidth, which efficiently handle the parallel matrix operations and tensor computations central to deep learning.

Furthermore, our cross-framework evaluation identifies critical software-level optimizations. Specifically, TensorFlow’s kernel-fusion techniques resulted in a 15\% reduction in inference latency compared to PyTorch, highlighting that architectural efficiency is as much a product of software engineering as it is of hardware design. While GPUs exhibit higher memory consumption due to specialized management techniques like coalesced access, this tradeoff is justified by superior performance-per-watt and throughput in memory-intensive tasks.

Beyond performance benchmarking, this study highlights a growing tension between the computational demands of modern AI models and the limited availability of energy-efficient hardware. Our results support the argument that democratized, shared access to GPU resources is essential for sustaining AI innovation across diverse research communities. Without such approaches, advances in AI risk are becoming increasingly centralized and inaccessible.

Projecting into the future, our trend analysis based on two decades of hardware data indicates that GPU memory capacity is on a non-linear upward trajectory. As deep learning models continue to scale in parameter count and complexity, the democratization of high-performance computing through cloud-based GPU resources will remain vital for global AI development. Our findings conclude that continued innovation in both hardware parallelism and framework-specific optimizations is essential to support the next generation of scalable and efficient AI applications. 
\bibliographystyle{IEEEtran}
\bibliography{ref}

@article{b1,
  author  = {Chandran, N. and Gangodkar, D. and Mittal, A.},
  title   = {A Review on GPU Programming Strategies and Recent Trends in GPU Computing},
  journal = {Journal of Graphic Era University},
  year    = {2018},
  volume  = {6},
  number  = {2},
  pages   = {207--223}
}

@article{b3,
  author  = {Nickolls, J. and Dally, W. J.},
  title   = {The GPU computing era},
  journal = {IEEE micro},
  year    = {2010},
  volume  = {30},
  number  = {2},
  pages   = {56--69}
}

@article{b4,
  author  = {Coelho, I. M. and others},
  title   = {A GPU deep learning metaheuristic based model for time series forecasting},
  journal = {Applied Energy},
  year    = {2017},
  volume  = {201},
  pages   = {412--418}
}

@article{b5,
  author  = {Owens, J. D. and others},
  title   = {GPU computing},
  journal = {Proceedings of the IEEE},
  year    = {2008},
  volume  = {96},
  number  = {5},
  pages   = {879--899}
}

@article{v10,
  author = {Simonyan, K. and Zisserman, A.},
  title  = {Very deep convolutional networks for large-scale image recognition},
  journal = {arXiv preprint arXiv:1409.1556},
  year   = {2014}
}

@inproceedings{v11,
  author    = {He, K. and others},
  title     = {Deep residual learning for image recognition},
  booktitle = {Proceedings of the IEEE conference on computer vision and pattern recognition},
  year      = {2016},
  pages     = {770--778}
}

@incollection{b6,
  author    = {Chowdhary, K. R.},
  title     = {Natural language processing},
  booktitle = {Fundamentals of Artificial Intelligence},
  publisher = {Springer},
  year      = {2020},
  pages     = {603--649},
  address   = {New Delhi}
}

@book{b8,
  author    = {Osman, A. A. M.},
  title     = {GPU Computing Taxonomy},
  publisher = {Recent Progress in Parallel and Distributed Computing},
  year      = {2017},
  volume    = {45}
}

@book{b9,
  title={Programming on Parallel Machines},
  author={Matloff, Norman},
  year={2011},
  publisher={University of California, Davis}
}

@article{bb01,
  title={Modelling using polynomial regression},
  author={Ostertagov{\'a}, Eva},
  journal={Procedia Engineering},
  volume={48},
  pages={500--506},
  year={2012},
  publisher={Elsevier}
}

@article{bb02,
  title={Algorithms of parametric estimation of polynomial trend models of time series on discrete transforms},
  author={Ismagilov, II and Khasanova, SF},
  journal={Academy of Strategic Management Journal},
  volume={15},
  pages={20},
  year={2016}
}

@mastersthesis{v9,
  title={Learning multiple layers of features from tiny images},
  author={Krizhevsky, Alex},
  year={2009},
  school={University of Toronto}
}

@inproceedings{b12,
    author = {Misic, Marko and Durdvic, Dorde and Tomasevic, Milo},
    year = {2012},
    month = {01},
    pages = {289-294},
    title = {Evolution and trends in GPU computing},
    isbn = {978-1-4673-2577-6}
}

@article{b13,
  title={Accelerating the XGBoost algorithm using GPU computing},
  author={Mitchell, Rory and Frank, Eibe},
  journal={PeerJ Computer Science},
  volume={3},
  pages={e127},
  year={2017},
  publisher={PeerJ Inc.}
}

@inproceedings{b14,
  title={Using GPUs for machine learning algorithms},
  author={Steinkraus, Dave and Buck, Ian and Simard, Patrice Y},
  booktitle={Eighth International Conference on Document Analysis and Recognition (ICDAR'05)},
  pages={1115--1120},
  year={2005},
  organization={IEEE}
}

@article{b16,
  title={A survey on techniques for cooperative CPU-GPU computing},
  author={Raju, K and Chiplunkar, Niranjan N},
  journal={Sustainable Computing: Informatics and Systems},
  volume={19},
  pages={72--85},
  year={2018},
  publisher={Elsevier}
}

@misc{b18,
  author = {Sumanaweera, Thilaka S. and Cherry, John W.},
  title = {Methods and systems for medical diagnostic and simulation with a graphics processing unit},
  year = {2006},
  howpublished = {U.S. Patent No. 7,119,810},
  month = {Oct},
  assignee = {Siemens Medical Solutions USA, Inc.}
}

@inproceedings{b19,
  title={GPU Accelerated Industrial Data Analysis in Private Cloud Environment},
  author={Artem N. Sisyukov and Olga Yulmetova and V. A. Kuznecov},
  booktitle={2019 IEEE Conference of Russian Young Researchers in Electrical and Electronic Engineering (EIConRus)},
  year={2019},
  pages={348-352},
  doi={10.1109/EICONRUS.2019.8656751}
}

@inproceedings{b21,
  title={Analyzing Machine Learning Workloads Using a Detailed GPU Simulator},
  author={Jonathan Lew and Deval Shah and others},
  booktitle={2019 IEEE International Symposium on Performance Analysis of Systems and Software (ISPASS)},
  year={2019},
  doi={10.1109/ISPASS.2019.00028}
}

@inproceedings{b23,
  author = {Dutta, Bishwajit and Adhinarayanan, Vignesh and Feng, Wu-chun},
  title = "{GPU Power Prediction via Ensemble Machine Learning for DVFS Space Exploration}",
  booktitle = {2018 ACM International Conference on Computing Frontiers (CF)},
  address = {Ischia, Italy},
  month = {May},
  year = {2018},
}

@phdthesis{b004,
  title={Reevaluating the Role of GPUs in Artificial Intelligence: A Comprehensive Analysis and Sustainable Future Directions.},
  author={Amin, Al},
  year={2012},
  school={University of Nevada}
}

@misc{b37,
  title={Cifar-10 (Canadian Institute for Advanced Research)},
  author={Alex Krizhevsky and Vinod Nair and Geoffrey Hinton},
  year={2010},
  howpublished={\url{http://www.cs.toronto.edu/kriz/cifar.html}},
  note={5(4), 1}
}

@article{b39,
  title={Benchmarking TPU, GPU, and CPU platforms for deep learning},
  author={Yuan Yao Wang and Gu-Yeon Wei and David Brooks},
  journal={arXiv preprint arXiv:1907.10701},
  year={2019}
}

@inproceedings{b41,
  title={Where is the data? Why you cannot debate CPU vs. GPU performance without the answer},
  author={C. Gregg and K. Hazelwood},
  booktitle={IEEE International Symposium on Performance Analysis of Systems and Software (ISPASS)},
  year={2011},
  pages={134-144},
  month={April}
}

@article{b43,
  title={Summarizing CPU and GPU design trends with product data},
  author={Yijin Sun and N. B. Agostini and S. Dong and David Kaeli},
  journal={arXiv preprint arXiv:1911.11313},
  year={2019}
}

@inproceedings{b47,
  title={Bridging the semantic gaps of GPU acceleration for scale-out CNN-based big data processing: Think big, see small},
  author={Song, Mingcong and Hu, Yang and Xu, Yunlong and Li, Chao and Chen, Huixiang and Yuan, Jingling and Li, Tao},
  booktitle={Proceedings of the 2016 International Conference on Parallel Architectures and Compilation},
  pages={315--326},
  year={2016}
}

@article{b48,
  title={Adabatch: Adaptive batch sizes for training deep neural networks},
  author={Devarakonda, Aditya and Naumov, Maxim and Garland, Michael},
  journal={arXiv preprint arXiv:1712.02029},
  year={2017}
}

@article{b49,
  title={MLOps Scaling Machine Learning Lifecycle in an Industrial Setting},
  author={Zhao, Yizhen and Belloum, Adam SZ and Zhao, Zhiming and others},
  journal={International Journal of Industrial and Manufacturing Engineering},
  volume={16},
  number={5},
  pages={138--148},
  year={2022}
}

@article{b50,
  title={Toward Understanding and Testing Deep Learning Information Flow in Deep Learning-Based Android Apps},
  author={Zhang, Jie and Guo, Qianyu and Zhang, Tieyi and Feng, Zhiyong and Li, Xiaohong},
  journal={International Journal of Computer and Systems Engineering},
  volume={17},
  number={3},
  pages={171--179},
  year={2023}
}

@article{b51,
  title={Adopting Cloud-Based Techniques to Reduce Energy Consumption: Toward a Greener Cloud},
  author={Achar, Sandesh},
  journal={International Journal of Computer and Information Engineering},
  volume={17},
  number={1},
  pages={70--77},
  year={2023}
}

@article{b52,
  title={Towards sustainable AI: a comprehensive framework for Green AI},
  author={Tabbakh, Abdulaziz and Al Amin, Lisan and Islam, Mahbubul and Mahmud, GM Iqbal and Chowdhury, Imranul Kabir and Mukta, Md Saddam Hossain},
  journal={Discover Sustainability},
  volume={5},
  number={1},
  pages={408},
  year={2024},
  publisher={Springer}
}

@inproceedings{b54,
  title={Microsoft coco: Common objects in context},
  author={Lin, Tsung-Yi and Maire, Michael and Belongie, Serge and Hays, James and Perona, Pietro and Ramanan, Deva and Doll{\'a}r, Piotr and Zitnick, C Lawrence},
  booktitle={Computer Vision--ECCV 2014: 13th European Conference, Zurich, Switzerland, September 6-12, 2014, Proceedings, Part V 13},
  pages={740--755},
  year={2014},
  organization={Springer}
}

@article{b55,
  title={Imagenet classification with deep convolutional neural networks},
  author={Krizhevsky, Alex and Sutskever, Ilya and Hinton, Geoffrey E},
  journal={Advances in neural information processing systems},
  volume={25},
  year={2012}
}

@INPROCEEDINGS{b56,
  author={Wang, Yuxin and Wang, Qiang and Shi, Shaohuai and He, Xin and Tang, Zhenheng and Zhao, Kaiyong and Chu, Xiaowen},
  booktitle={2020 20th IEEE/ACM International Symposium on Cluster, Cloud and Internet Computing (CCGRID)}, 
  title={Benchmarking the Performance and Energy Efficiency of AI Accelerators for AI Training}, 
  year={2020},
  volume={},
  number={},
  pages={744-751},
  }

@article{b57,
  title={A survey on platforms for big data analytics},
  author={Singh, Dilpreet and Reddy, Chandan K},
  journal={Journal of big data},
  volume={2},
  pages={1--20},
  year={2015},
  publisher={Springer}
}

@ARTICLE{hopper_perf,
  author={Choquette, Jack},
  journal={IEEE Micro}, 
  title={NVIDIA Hopper H100 GPU: Scaling Performance}, 
  year={2023},
  volume={43},
  number={3},
  pages={9-17},
  doi={10.1109/MM.2023.3256796}
}

@article{gpu_H100,
  title={{NVIDIA H100 Tensor Core GPU Architecture}},
  author={NVIDIA, Hopper},
  journal={NVIDIA, Santa Clara, Calif, USA},
  year={2022}
}

@phdthesis{phd_thesis,
  title={Efficient Memory Coherence and Consistency Support for Enabling Data Sharing in GPUs},
  author={Tabbakh, Abdulaziz},
  year={2018},
  school={University of Southern California}
}

@misc{llmssurvey,
      title={Large Language Models: A Survey}, 
      author={Shervin Minaee and Tomas Mikolov and Narjes Nikzad and Meysam Chenaghlu and Richard Socher and Xavier Amatriain and Jianfeng Gao},
      year={2025},
      eprint={2402.06196},
      archivePrefix={arXiv},
      primaryClass={cs.CL},
      url={https://arxiv.org/abs/2402.06196}, 
}

@INPROCEEDINGS{aimlcompute,
  author={Sevilla, Jaime and Heim, Lennart and Ho, Anson and Besiroglu, Tamay and Hobbhahn, Marius and Villalobos, Pablo},
  booktitle={2022 International Joint Conference on Neural Networks (IJCNN)}, 
  title={Compute Trends Across Three Eras of Machine Learning}, 
  year={2022},
  volume={},
  number={},
  pages={1-8},
  doi={10.1109/IJCNN55064.2022.9891914}}

\begin{IEEEbiography}[{\includegraphics[width=1in,height=1.25in,clip,keepaspectratio]{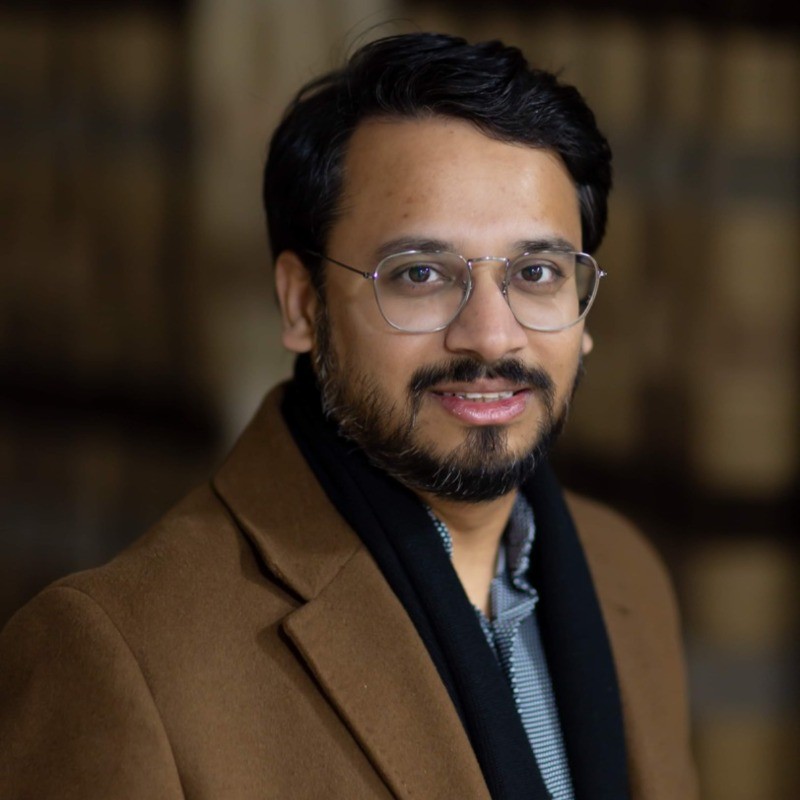}}]{Lisan Al Amin} is a Senior Data Analyst and Data Scientist with over eight years of experience in healthcare analytics, machine learning, and AI-driven decision systems. He is the founder and CEO of Cognitive Links LLC, where he leads innovation in educational and enterprise AI solutions. He currently serves as a federal contractor with the U.S. Health Resources and Services Administration (HRSA) through REI Systems, developing advanced statistical models and AI-powered dashboards to assess clinical quality, reduce health disparities, and enhance telehealth services. He is also an Adjunct Faculty member at the Washington University of Science and Technology (WUST) and a mentor for aspiring data scientists at Flatiron School. Lisan holds dual master’s degrees in Information Systems and Computer Science and is currently pursuing a Ph.D. in Machine Learning at the University of Maryland. His research interests include quantum-enhanced AI, healthcare analytics, and sustainable machine learning. He has published in leading journals and conferences and actively contributes as a peer reviewer and presenter.

\end{IEEEbiography}

\begin{IEEEbiography}[{\includegraphics[width=1in,height=1.25in,clip,keepaspectratio]{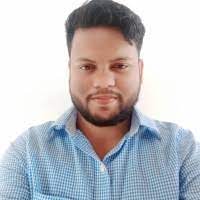}}]{Md. Ismail Hossain} received his B.Sc. degree in Computer Science and Engineering from North South University (NSU), Bangladesh. He is currently a researcher at the Apurba-NSU R\&D Lab and a Fatima Fellow, selected for his work on model efficiency and knowledge distillation in deep learning. His research spans machine learning, model compression (including pruning and distillation), and domain adaptation. He has collaborated on interdisciplinary projects with academic institutions and NGOs and has served as a reviewer for major AI conferences such as NeurIPS and ICML.
\end{IEEEbiography}



\begin{IEEEbiography}[{\includegraphics[width=1in,height=1.25in,clip,keepaspectratio]{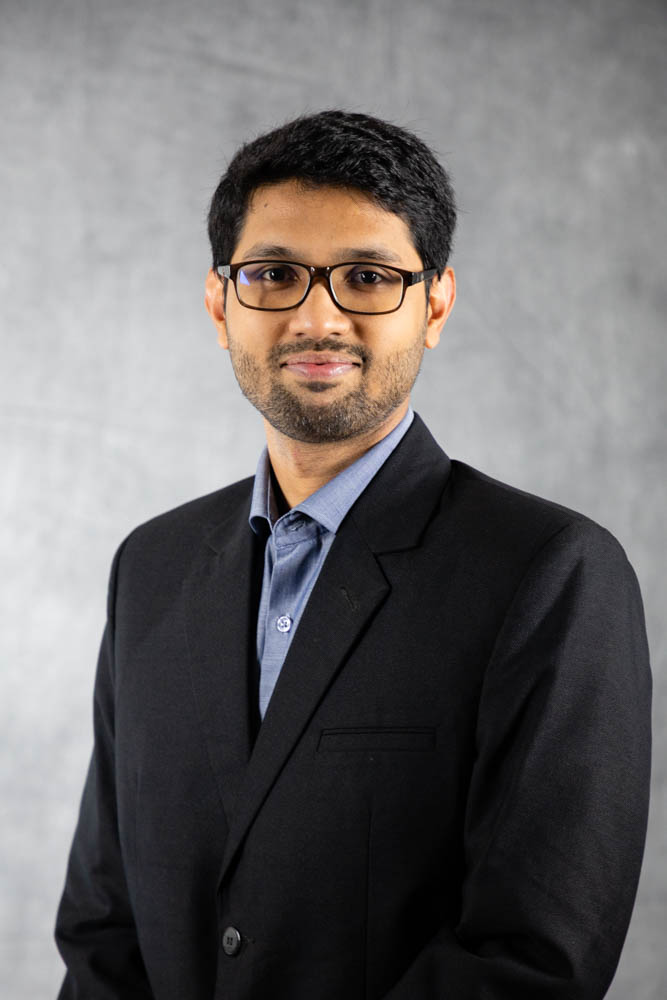}}]{Rupak Kumar Das} received the B.S. from Khulna University of Eng and Tech, Bangladesh, and M.S. degrees in Computer Science from the University of Minnesota Duluth.
He is currently a third-year Ph.D. student in Informatics at Pennsylvania State University, University Park, PA, USA.
He is also enrolled in an NSF-funded Linguistic Diversity (LinDiv) traineeship program. 
His research interests include Natural Language Processing, Machine Learning, and Explainable AI.
\end{IEEEbiography}

\begin{IEEEbiography}[{\includegraphics[width=1in,height=1.25in,clip,keepaspectratio]{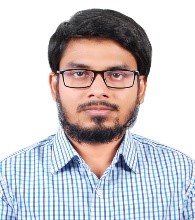}}]{Mahbubul Islam } received the B.Sc. in Electrical and Electronics Engineering (EEE) from Ahsanullah University of Science and Technology and M.Sc. in Computer Science and Engineering (CSE) in United International University, Dhaka, Bangladesh. He is serving as Network Engineer (Assistant Manager) under the Gateway Operations department in Summit Communications, Bangladesh and also as a Senior Research Associate, Data Science Track (Remote) in Cognitive Links, LLC, Maryland, USA. He is an IEEE Graduate student member (GSMIEE) and a Member of IEB (MIEB). His research interests include Human Computer Interaction, Natural Language Processing, Applied Machine learning, and Data Mining.

\end{IEEEbiography}

\begin{IEEEbiography}[{\includegraphics[width=1in,height=1.25in,clip,keepaspectratio]{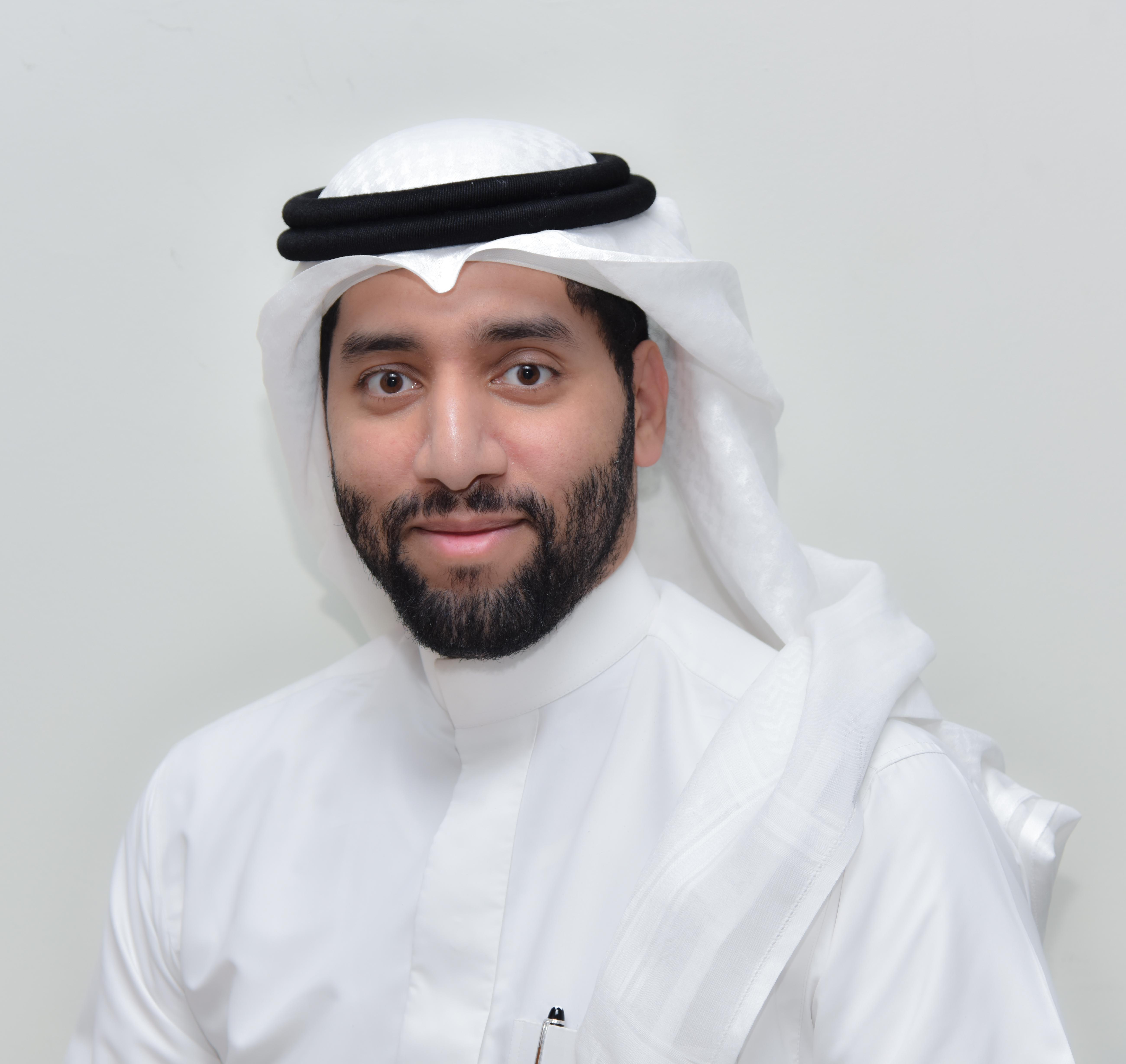}}]{Abdulaziz Tabbakh} received the B.S. and M.S. degrees in computer engineering from
King Fahd University of Petroleum and Minerals (KFUPM), Saudi Arabia, in 2007 and 2011, respectively. He received the Ph.D. degree in electrical and computer engineering from the University of Southern California (USC), Los Angeles, CA, USA, in 2018. 
Since 2018, he has been an Assistant Professor in the
Computer Engineering Department at King Fahd University of Petroleum and Minerals (KFUPM), Saudi Arabia.
His research interests include high-performance computing, computer architecture, GPU architecture and hardware acceleration for AI/ML applications.

\end{IEEEbiography}
\EOD

\end{document}